\documentclass[12pt,a4paper]{article}
\usepackage{amssymb,amsfonts,amsmath}
\usepackage{color}
\usepackage{epsfig, euscript, graphics}

\begin{document}

\title{Contextuality, Complementarity, Signaling, and Bell tests}

\author{Andrei Khrennikov\\ 
Linnaeus University, International Center for Mathematical Modeling\\  in Physics and Cognitive Sciences
 V\"axj\"o, SE-351 95, Sweden}

\maketitle
 
\abstract{This is a review devoted to the complementarity-contextuality  interplay  with connection to the Bell inequalities.  Starting discussion with complementarity, we  point out to contextuality as its seed. {\it Bohr-contextuality}  is dependence of observable's outcome on the experimental context, on system-apparatus  interaction. Probabilistically, complementarity means that  the {\it  joint probability distribution} (JPD)  does not exist. Instead of the JPD, one has to operate with contextual probabilities. The Bell inequalities are interpreted as the statistical  tests of contextuality and, hence, incompatibility. For context dependent probabilities, these inequalities may be violated.  We stress that contextuality tested by the Bell inequalities is so called {\it joint measurement contextualit}y (JMC), the special case of Bohr's contextuality. Then, we examine the role of signaling  (marginal inconsistency). In QM, signaling can be considered as an experimental  artifact.  However, often experimental  data has signaling patterns. We discuss possible  sources of signaling; for example, dependence of the state preparation on measurement  settings.  In principle, one can extract the measure of ``pure contextuality''  from  data shadowed by signaling.  This theory known as {\it Contextuality by Default} (CbD). It leads to inequalities with the additional term quantifying signaling, Bell-Dzhafarov-Kujala inequalities.}

{\bf keywords:} contextuality; complementarity; Bell inequalities; quantum nonlocality; joint probability distribution; 
V\"axj\"o model for contextual probability; signaling; Contextuality by Default 

\section{Introduction}

This is a review devoted to the interplay of notions of contextuality and complementarity as the interpretational basis of the violation of the Bell inequalities \cite{Bell}-\cite{BellSP}. We set essential efforts  to clarify and logically structure Bohr's views  \cite{BR0} on contextuality  and contextuality's crucial role in the derivation of the complementarity principle \cite{KHR7}-\cite{KHR7c}(see also \cite{JaegerC1, JaegerC2}). In fact, in Bohr's writings these two notions are really inseparable. We recommend to the reader the books of Plotnitsky and Jaeger \cite{PL1}-\cite{Ja2} clarifying Bohr's views on complementarity and contextuality. Bohr did not use  the notion of contextuality. He wrote about experimental conditions. But in the modern terminology he appealed to contextuality of quantum measurements.  We remark that at the beginning Bell neither used this terminology. This notion was invented in QM by Beltrametti  and Cassinelli \cite{Beltrametti}.

In philosophic terms Bohr's contextuality means rejection of {\it ``naive realism''}; by Bohr the outcomes of quantum measurements cannot be treated as the objective properties of a system under observation.  These values cannot be assigned to a system before a  measurement,
with exception of special system's states - the eigenstates of observables. However, we do not like to operate with the notion of realism including the EPR elements of reality. We leave this field for philosophers who have been working on it during the last two thousands years. Instead we will work with the notion of Bohr's contextuality which is formulated in the heuristically clear physical terms - the interaction between a system and a measurement device. We would neither operate with the notion of {\it local realism}. I think that this is an ambiguous notion, but this is just  my personal viewpoint.  At least one has to split local realism into two components, realism and locality, and then analyze them separately. We will shortly discuss this notion and its components 
in appendix A.         

In this review we do not try to cover all approaches to contextuality; in particular, we do not discuss the Kochen-Specker theorem and the corresponding contextuality (see the recent review of Svozil \cite{Svozil} for the description of the diversity of the views on contextuality).  

 Starting with mentioning the Bohr principle of complementarity also known as ``wave-particle duality'' , we  analyze the notion of contextuality. The latter is understood very generally, as the irreducible dependence of observable's outcome on the experimental context. Thus, the outcomes of quantum observables are not the objective properties of systems. They are generated in the complex process of interaction between a system and a measurement device.  In fact,  {\it ``Bohr-contextuality''} is the seed of complementarity, the existence of incompatible observables \cite{KHR7}-\cite{KHR7c}.  

In the probabilistic terms, incompatibility means that  JPD  does not exist. Instead of the JPD, one has to operate with a family of probability distributions depending on experimental contexts  as in the {\it the  V\"axj\"o model}  for contextual probability theory \cite{INT}-\cite{KHR5a}. This model generalizes the notion of conditional probability from classical probability  (CP) theory. In some cases the contextual probability update can be represented via the state update of the projection type represented in the 
complex Hilbert space \cite{KHR1}-\cite{KHR3a}, \cite{KHR5}, \cite{BNY1,BNY2}. And, of course, the probability update of quantum theory can be easily realized as update of contextual probability. The update machinery is formalized via introduction of special contexts corresponding to the outcomes of observables 
\cite{INT}-\cite{KHR5a}. 

We continue to analyze the probabilistic structure of QM by considering the Bell inequalities and concentrating on the CHSH-inequality \cite{CHSH} and the Fine theorem \cite{Fine}. This theorem connects Bell inequality with the existing of the JPD
for four observables involved in the Bohm-Bell experiment, in fact the group of four separate experiments for 
the pairwise measurements for some pairs of these observables. We use the Fine theorem as the bridge to the 
contextual interpretation of the Bell type inequalities. For context dependent probabilities in the absence of JPD unifying them, these inequalities can be violated \cite{KHR5}.  We point out that contextuality tested by the Bell inequalities is so called {\it joint measurement contextuality} (JMC) \cite{Bell2} (and section \ref{JMC7})  -- the very special case of Bohr's contextuality. We stress that consideration of JMC is dominating within the quantum studies of contextuality. On one hand, this simplifies the picture; on the other hand, by reducing Bohr's contextuality to JMC people miss the general contextual perspective as it was established by Bohr at the very beginning of QM. 

Some authors even define contextuality directly as the violation of some Bell inequality (see, eg.,  \cite{Araujo} and references herein).  We call such type of contextuality {\it Bell contextuality.} However, Bell by himself invented contextuality \cite{Bell2} as JMC and then he pointed out that JMC can serve as a source of ``Bell contextuality''.

We remark that originally Bell explained the violation of his inequality by Einsteinian nonlocality \cite{EPR}, ``spooky action at a distance'' - Einstein's hype slogan. In article \cite{Bell2} Bell discussed contextuality in the JMC form in connection  with nonlocality (see also related papers of Gudder \cite{GD1}-\cite{GD3} and Shimony \cite{Sh1,Sh2}). However, JMC per se cannot clarify the origin of  Einsteinian nonlocality.  In Bell's discussion \cite{Bell2} JMC looks even more mystical than nonlocality. Consideration of JMC as the special case of Bohr contextuality and connecting it with incompatibility, demystifies JMC. And by highlighting the role of incompatibility, the debate on the meaning of the Bell type inequalities turns to the very basics of QM, to Bohr's complementarity principle and the existence of incompatible observables. {\it The Bell inequalities are interpreted as the special tests of contextuality and, hence, incompatibility} \cite{NL1, NL3}. Coupling contextuality-incompatibility is basic in our treatment of the Bell inequalities.  This review continues the line of articles -- {\it ``getting rid off nonlocality from quantum physics''} 
\cite{NL1}-\cite{KHR7b}  (see also \cite{Cetto0}-\cite{Cetto}).

We also examine {\it signaling} which may be better to call {\it marginal inconsistency} by following the line of research presented in articles of Adenier and Khrennikov   \cite{AD3}-\cite{AD2}. Typically its role in discussions on the Bell inequalities is not highlighted. In contrast to the majority of authors, we take very seriously complications related to the presence signaling patterns in experimental statistical data \cite{AD1}.  It must be noted that the terminology ``signaling'' is quite ambiguous, since in fact ``signaling''is defined not 
 in terms of signals propagating in physical space-time, but in purely probabilistic framework, as  non-coincidence of marginal probability distributions corresponding to join measurements of an observables $a$ with  other observables which are compatible with it. 

In QM, signaling can be considered as an experimental  artifact - theoretically there should be no signaling.  However, often experimental  data has signaling patterns which are  statistically non-negligible \cite{AD1}, \cite{ASP}-\cite{KHRR}. We discuss possible  sources of signaling, both in the theoretical and experimental frameworks. In particular, we point out to dependence of the state preparation procedure on settings of measurement devices as a signaling source (cf. \cite{ASP,Wei,Wei1}): the standard source state generation is supplemented  with additional state modification which is setting dependent. We emphasize that in the studies on interrelation between classical and quantum physics, signaling cannot be ignored. The presence of signaling in the experimental statistical data
 per se means that such data cannot be modeled within QM. So, in such a case there is no need to check whether some Bell inequality is violated or not.  In the presence of signaling approaching the high level of the violation of  e.g. the CHSH-inequality is totally meaningless. Even tremendous efforts to close all possible loopholes meaningless if data 
suffers of signaling.  

We remark that, as was recently found by Dzhafarov et al. \cite{DZ1}-\cite{DZ7}, one can extract the measure of pure contextuality even from statistical data shadowed by signaling.  This theory known as {\it Contextuality by Default} (CbD) is based on {\it coupling technique} of CP.  CbD with mathematical technique from CP leads to the Bell inequalities with the additional term quantifying the level of signaling,
we call such inequalities the {\it Bell-Dzhafarov-Kujala inequalities}  (BDK). In this review, we are concentrated on the CHSH-BDK inequality.  Generally, CbD can be considered as a part of the project on the CP-treatment of the Bell inequalities and contextuality. Another part of this project was presented in \cite{Avis}-\cite{KHR6a}, where {\it quantum probabilities were treated as classical conditional probabilities} with conditioning w.r.t. the selection of experimental settings (cf. with Koopman  \cite{Koopman}, Ballentine   \cite{BL0}, \cite{BL}-\cite{BL2}). This is the good place to mention the CP-based tomographic approach to QM which was developed by Vladimir Man'ko and coauthors \cite{MA1}-\cite{MA4}.   We also point out to articles \cite{DZ7} and \cite{KHRcomment} for a debate on the perspectives of the CP-use in contextual modeling (without direct connection with QM).

I also would like to  inform physicists that nowadays quantum theory, its methodology and mathematical formalism, are widely applied outside of physics, to cognition, psychology, decision making, social and political sciences, economics and finances (see, e.g., monographs \cite{QL0}-\cite{BAF} and references in them). I called this kind scientific research {\it quantum-like modeling} and this terminology was widely spread.  In particular, contextuality based on the quantum studies  attracted a lot of attention, especially in cognitive psychology and decision making, including the Bell tests \cite{QLX}, \cite{Conte}-\cite{BDZ}. One of the specialties of such studies is the presence of signaling patterns in statistical data collected in all experiments which were done up to now \cite{DZ3}. Here  the BDK-inequalities are especially useful \cite{CERV,BDZ}. 

In this review we discuss mainly the CHSH inequality. This is motivated by two reasons, experimental and theoretical ones. The basic 
of experiments were done for this inequality \cite{ASP,Wei,Hen,Sh,Wei0} (with some very important exceptions \cite{Gi0,Gi}, see also 
\cite{KHRR}). The mathematical structure of this inequality makes it possible to establish the straightforward coupling with incompatibility expressed mathematically in the form of commutators \cite{NL1} (section \ref{LLL6}). From my viewpoint, 
the original Bell inequality  derived under the assumption on the prefect correlations deserves more attention, both theoretically and experimentally; some steps in this direction were done in works \cite{KHR7a0}-\cite{LE}.

In this review we are concentrated only on the Bohr contextuality and its ``derivatives'', JMC and Bell contextuality. We neither discuss hidden variables theory. The latter may be surprising, since from the beginning the Bell  inequalities were derived in hidden variables framework. However, we treat these inequalities as statistical tests of incompatibility.
In the presence of incompatible observables, it is meaningless to discuss theories with hidden variables, at least theories in which hidden variables are straightforwardly connected with the outcomes of observables as was done by Bell and his followers. Already De Broglie pointed out that such theories have no physical meaning.

In principle, one can consider subquantum models, but variables of such models are only indirectly coupled to outcomes of quantum observables. The latter viewpoint was advertized by Schr\"odinger \cite{DA} who in turn followed the works fo Hertz  \cite{HER} and Boltzmann \cite{BZ1,BZ2} (see also \cite{KHRAN,KHRHERZ}). One of such subquantum theories was developed in the series of author's works on emergence of QM from classical random field theory  \cite{Beyond}.

\section{Preliminary discussion}

\subsection{Forgotten contribution of Bohr to contextuality theory}

Contextuality is one of the hottest topics of modern quantum physics, both theoretical and experimental.
During the recent 20 years, it was discussed in numerous papers published in top physical journals. Unfortunate of these discussions is that from the very beginning contextuality (JMC, section \ref{JMC7}) was coupled to the issue of nonlocality. It was Bell's intention in his analysis of the possible seeds of the violation of the Bell type inequalities \cite{Bell2}.

Surprisingly, Bell had never mentioned general contextuality which we call {\it ``Bohr contextuality''}. The latter has no straightforward coupling to the Bell inequalities; it is closely related to the notion of incompatibility of observables - 
the {\it Bohr principle of complementarity.} What is even more surprising that Shimony who was one of authorities in quantum foundations by commenting  \cite{Sh1,Sh2} Bell's article \cite{Bell2} had neither mentioned the Bohr principle of complementarity and its contextual dimension. 

One of the explanations for this astonishing situation in quantum foundations is that Bohr presented his ideas in a vague way; moreover, he often changed his vague formulations a few times at different occasions.  In this section we briefly present  Bohr's ideas about contextuality of quantum measurements and its role in his formulation of the complementarity principle (see 
\cite{KHR7}-\cite{KHR7c} for detailed presentations). Then, we move to the Bell inequalities. This pathway towards these inequalities (i.e., via Bohr's contextuality-complementarity) highlights the role of incompatibility of quantum observables in the Bell framework and gives the possibility to operate with the Bell inequalities without mentioning the ambiguous notion of quantum nonlocality (spooky action at a distance).     

\subsection{What does contextuality mean?}

In this situation when so many researchers write and speak about quantum contextuality, one should be sure that this notion is well defined and its physical interpretation is clear and well known. In fact, before started to think about the meaning 
of contextuality, I was completely sure in this. Strangely enough, I  was not able to create a consistent picture. 
And I was really shocked when by visiting the institute of Atom Physics in Vienna and having conversation with  
 Rauch and Hasegawa, I found that they are also disappointed. They asked me about the contextuality meaning. And they performed the brilliant experiments  \cite{Rauch0,Rauch1} to test contextuality in the framework of neuron interferometry. They had a vague picture of what was tested and what is the physical meaning of their experimental results!

Then, in Stockholm by being in the PhD defense  jury  of one student who was supervised by  prof. Bengtsson (let call her Alice), I asked Alice about  the physical meaning of contextuality. (Her thesis was about it.) Alice  answered that she has no idea about 
the physical interpretation of advanced mathematical results obtained in her thesis. Generally I like discussions. To stimulate a debate,   I told that Rauch and Hasegawa
had the strange idea that contextuality is just noncommutativity, a sort of the order effect in the sequential measurements (this was the final output of our discussions in Vienna). Unfortunately, in Stockholm the discussion  quickly finished with the conclusion that the question is interesting, but not for the PhD-defense.

\subsection{Jump from contextuality to Bell inequalities}
\label{JMC7}

Typically by writing a paper about contextuality in QM one starts by referring to this notion as {\it joint measurement contextuality}
(JMC):
\medskip
 {\it dependence of the outcomes of some observable $a$ on its joint measurement with another observable $b.$ } 
\medskip 
We note that this definition is countefactual and cannot be used in the experimental framework.

 Nevertheless, the ``universal contextuality writer'' is not disappointed by this situation and he immediately jumps to the Bell inequalities which are treated
 as {\it noncontextual inequalities} (see,e.g., \cite{Araujo}). Moreover, contextuality is often  identified with the violation of the Bell inequalities - Bell contextuality in our terminology. This identification shadows the problem of the physical meaning of contextuality. One jumps  from the problem of understanding to calculation of a numerical quantity, the degree of the violation of some  Bell inequality.
Such inequalities are numerous. And they can be tested in different experimental situations and  generate the permanent flow of highly recognized papers.   

I suggested the following critical illustration to this strategy (contextuality = violation of the Bell inequalities) \cite{KHR7d}.  
Consider the notion of a  {\it random sequence.} Theory of randomness is the result of the intensive research (Mises, Church, Kolmogorov, Solomovov, Chatin,  Martin-L\"of ; see,e.g., the first part of my book  \cite{KHRWS}). This theoretical basis led to elaboration of the variety of randomness tests which are used to check whether some sequence of outputs of physical or digital random generator is random. 
But, in fact, it is possible to check only pseudo-randomness. The universal test of randomness, although exist, but the proof of its existence is nonconstructive and this test cannot be applied to the concrete sequence of outcomes.  

In applications the NIST test (a batch of tests for randomness) is the most widely used. So, in theory of randomness  we also use tests, but beyond them there is the well developed theory of randomness. In particular, this leads to understanding that even if a sequence $x$ passed the NIST test, this does not imply that it is random. In principle, there can be found another test such that $x$ would not pass it. The latter would not be a surprise.

In contrast to the above illustration, in QM contextuality is per definition the violation of some noncontextual (Bell) inequality (at least for some authors). Hence, the theoretical notion is identified with the Bell test; in fact, the batch of the 
tests corresponding to different Bell inequalities. (The Bell test for classicality plays the role of the NIST test for randomness).  This is really bad! Not only from the theoretical viewpoint, but even from the practical one. As was mentioned, by working with randomness people understand well that even passing the NIST test does not guarantee randomness.
In QM, passing the Bell test is per definition is equivalent to contextuality. This is wrong strategy which led to skews  in handling quantum contextuality.

\subsection{Signaling and other anomalies in data} 

The first signs that addiction to one concrete test of contextuality (Bell inequalities)  may lead to the wrong conclusions
were observed by Adenier and Khrennikov \cite{AD3}-\cite{AD2}. 
Adenier was working on the translation of the PhD thesis of Alain Aspect (due to the joint agreement with prof. Aspect and Springer)
and he pointed out to me that he found some strange anomalies in Aspect's data  \cite{ASP}.
One of them was signaling. i.e., dependence of detection probability on one side (Bob's lab) on the selection of an experimental setting 
on another side (Alice's lab). 

Then, we found signaling in the data from the famous  Weihs experiment closing the nonlocality loophole \cite{Wei}. Our publications \cite{AD3}-\cite{AD2} attracted attention to the problem of signaling in data collected in quantum experiments. 
Slowly people started to understand that experimenter cannot be happy by just getting higher degree of the violation of say the CHSH-inequality, with higher confidence. Often this implied the increase of the degree of signaling. Experimenters started to check the hypothesis of signaling in data \cite{Gi,Sh}. Unfortunately, our message was ignored by some experimenters, e.g., the data from the ``the first loophole free experiment'' \cite{Hen} demonstrated statistically significant signaling. 

\medskip

{\it Any Bell test should be combined with the test of experimental statistical data on signaling.}

\medskip

We pointed out that signaling was not the only problem in Aspect's data. As he noted in his thesis \cite{ASP}, the data contains 
``anomalies'' of the following type. Although the CHSH-combination of correlations violates the CHSH-inequality, the correlation for the concrete pair of angles $\theta_1,  \phi,$ as the function of these angles, does not match the theoretical prediction of QM, the graph of the experimental data differs essentially from the theoretical $\cos$-graph. Our attempts to discuss this problem with other experimenters generated only replies that ``we do not have such anomalies in our data''.\footnote{One of the problems in testing the Bell inequalities is the absence of openly approachable experimental data. I and Adenier struggled a lot to create the open database. But, in spite the positive responses, it seems that it was not created. I even wrote a paper ``Unuploaded experiments have no result"'
\cite{UNUP}.}

\subsection{V\"axj\"o model: Contextuality-complementarity and probability}
\label{VBM} 
    
In the probabilistic terms complementarity, incompatibility of observables, means that  
their {\it  joint probability distribution} (JPD) does not exist. Instead of the JPD, one has to operate with  context-dependent family of probability spaces - {\it the  V\"axj\"o probability model} \cite{INT}-\cite{KHR5a}: 
$$
{\cal M}_{{\cal  Z}}=({\cal P}_{C}, C \in {\cal  Z}),
$$ 
where ${\cal  Z}$ is a family of contexts and, for each $C \in {\cal  Z},$  
$$
{\cal P}_{C} =(\Omega_C, {\cal F}_C, P_C)
$$ 
is Kolmogorov probability space (appendix B). Here $\Omega_C$ is a sample space, ${\cal F}_C$ is a $\sigma$-algebra of subsets of $\Omega_C$ (events), 
and $P_C$ is a probability measure on ${\cal F}_C.$  
All these structures depend on context $C.$ To develop a fruitful theory,   ${\cal  Z}$ must satisfy to some conditions on 
inter-relation between contexts. THese conditions give the possibility to create an analog of the CP calculus of conditional 
probabilities. 

  In CP the points of $\Omega_C$ represent elementary events, the most simple events which can happen within context $C.$ Although these events are elementary, their structure can be complex and include the events corresponding to appearance of  some parameters (``hidden variables'') for a system under observation and measurement devices, times of detection and so on.   

Observables are given by random variables on contextually-labeled probability spaces, measurable functions,
$a_C:  \Omega_C \to \mathbb{R}.$ The same semantically defined observable $a$ is represented by a family of random variables 
$(a_C, C \in  {\cal  Z}_a),$ where ${\cal  Z}_a$ is the family of contexts for which the $a$-observable can be measured.
In ${\cal M}_{{\cal  Z}}$ averages and correlations  are also labeled by contexts,
\begin{equation}
\label{Ctyq}
\langle a\rangle_C= E[a_C|P_C]= \int_{\Omega_C}   a_C(\omega)  dP_C(\omega) = \int_{\mathbb{R}}   x  dP_{a|C}(x), 
\end{equation} 
\begin{equation}
\label{Cty1q}
\langle a b \rangle_C= E[a_C b_C|P_C]=\int_{\Omega_C}   a_C(\omega)  b_C(\omega) dP_C(\omega) = \int_{\mathbb{R^2}}   x y  d P_{a,b|C}(x,y), 
\end{equation}
where $P_{a|C}$ is the probability distribution of $a_c$ and $P_{a,b|C}$ is the JPD of the pair of random variables $(a_C,b_C).$ 
In (\ref{Ctyq}) $C\in {\cal  Z}_a$ and in (\ref{Cty1q}) $C\in {\cal  Z}_{a,b}= {\cal  Z}_a \cap {\cal  Z}_b.$ Since in context 
${\cal  Z}_{a,b}$ both observables $a$ and $b$ are represented by random variables, namely, by  $a_C$ and $b_C,$ it is natural to assume
that in this context both observables can be measured and the measure-theoretic JPD  $P_{a,b|C}$ represents mathematically the JPD for joint measurements of the pair of observables $(a,b).$ 

In further sections, we  analyze the probabilistic structure of QM by considering the Bell inequalities and concentrating on the CHSH-inequality \cite{CHSH} and the Fine theorem \cite{Fine}.

\subsection{Summary of preliminary discussion}

We can conclude the discussion with a few statements:
\begin{itemize}
\item The theoretical definition of contextuality as JMC suffers of appealing to conterfactuals.
\item Identification of contextuality with the violation of the Bell inequalities is not justified, 
neither physically nor mathematically (in the last case such an approach does not match the mathematical tradition).
\item The Bell tests have to accompanied with test on signaling.
\item ``Unuploaded to internet experiments have no results'' \cite{UNUP}. 
\item Probabilistically contextuality-complementarity is described by contextual probability (as by  
the V\"axj\"o model).
\end{itemize}

\section{Rethinking Bohr's ideas}

This section is devoted to rethinking of Bohr's foundational works in terms of contextuality. I spent a few years for reading Bohr and rethinking his often fuzzy formulations. 

\subsection{Bohr Contextuality}

The crucial question is about the physical meaning of contextuality; without answering to it, JMC (even by ignoring counterfactuality) is mystical, especially for spatially separated systems. Even spooky action at a distance is welcome - to resolve this mystery. 

In series of my papers \cite{KHR7}-\cite{KHR7c} the physical meaning of contextuality was clarified through referring to the Bohr's complementarity principle.    Typically this principle is reduced to wave-particle duality. (In fact, Bohr had never used the  latter terminology.) However, Bohr's formulation of the complementarity principle is essentially 
deeper. Complementarity  is not postulated; for Bohr, it is the natural consequence 
of the irreducible dependence of observable's outcome on the experimental context. Thus, the outcomes of quantum observables  are generated in the complex process of the interaction of a system and a measurement device \cite{BR0} (see also \cite{KHR7bb}, \cite{PLA}). 
This dependence on the complex of experimental conditions is nothing else than a form of contextuality, {\it Bohr-contextuality}
(section \ref{BPCC}).  We remark that JMC is its special case. But, in contrast to JMC, the physical interpretation of Bohr-contextuality is transparent - dependence of results of measurements on experimental contexts. And it does not involve the use of conterfactuals.     

Such contextuality  is the seed of complementarity, the existence of incompatible observables.  
(We recall that observables are incompatible  if they cannot be measured jointly.) Moreover, contextuality without incompatibility 
loses its value. 

\medskip

{\it If all observables were compatible, then they might be jointly measured in a single experimental context and multicontextual consideration would be meaningless.}   

\medskip

One can go in deeper foundations of QM and ask: 

\medskip

{\it Why is dependence on experimental context (system-apparatus interaction) is irreducible?}  

\medskip

Bohr's answer is that irreducibility is due to the existence of {\it indivisible quantum of action} given by the Planck constant
(see my article \cite{KHR7a,KHR7b} for discussion and references).

\subsection{Bohr's  Principle of Contextuality-complementarity}
\label{BPCC}

The Bohr principle of complementarity \cite{BR0} is typically presented as wave-particle duality, incompatibility of the position and 
momentum observables. The latter means the impossibility of their joint measurement. We remark that Bohr started with the problem of incompatibility of these observables by discussing the two slit experiment. In this experiment position  represented by ``which slit?'' observable and momentum is determined the detection dot on the registration screen. (This screen is covered by photo-emulsion and placed on some distance beyond the screen with two slits.) Later Bohr extended the wave-particle duality to arbitrary observables which cannot be jointly measured and formulated the principle of complementarity. He justified  this principle by emphasizing  contextuality of quantum measurements.  The Bohr's viewpoint on contextuality was wider than in the modern discussion on quantum contextuality related to the Bell inequality. The later is contextuality of joint measurement with a compatible observable (section \ref{JMC7}).   

In 1949,  Bohr \cite{BR0} presented the essence of complementarity in the following widely citing statement: 

\medskip

{\it ``This crucial point ...  implies the impossibility of any sharp separation between the behaviour of atomic objects and the interaction with the measuring instruments which serve to define the conditions under which the phenomena appear. In fact, the individuality of the typical quantum effects finds its proper expression in the circumstance that any attempt of subdividing the phenomena will demand a change in the experimental arrangement introducing new possibilities of interaction between objects and measuring instruments which in principle cannot be controlled. Consequently, evidence obtained under different experimental conditions cannot be comprehended within a single picture, but must be regarded as complementary in the sense that only the totality of the phenomena exhausts the possible information about the objects.''}  

\medskip

In short, Bohr's way to the complementarity principle, the claim on the existence of incompatible quantum observables, can be presented 
as the following chain of reasoning  \cite{KHR7}-\cite{KHR7c}: 
\begin{itemize}
 \item {\bf CONT1} An outcome of  any observable   is composed of the contributions of a system and a measurement device.\footnote{So, the values of an observable $a$ are not the objective properties of the systems.  They are created in the process of the complex interaction between the systems prepared for  measurements and the apparatus used for measurement of $a.$}
\item {\bf CONT2}  The whole experimental context has to be taken into account.
\item  {\bf INCOMP1}  There is no reason to expect that all experimental contexts can be combined with each other. 
\item {\bf INCOMP2} Therefore one cannot expect that all observables can be measured jointly.
\item {\bf INCOMP3} There can exist incompatible observables.  
\end{itemize}

The statements {\bf CONT1} + {\bf CONT2} and {\bf INCOMP1}+ {\bf INCOMP12}+ {\bf INCOMP3} compose the contextual and incompatibility parts of Bohr's reasoning.

Therefore it is more natural to speak about two Bohr's principles:
\begin{itemize}
\item {\it Contextuality Principle}.
\item {\it Complementarity Principle}.
\end{itemize}
And the second principle is a consequence of the first one. So, contextuality (understood in the Bohr's sense)  is the seed of complementarity. We can unify these two principles and speak about the {\bf Contextuality-Complementarity Principle}.  
Unfortunately, the contextual dimension of Bohr's complementarity is typically missing in the discussions on quantum foundations.
By speaking about the wave-particle duality one typically miss that the wave and particle properties of a system cannot be merged in a single 
experimental framework, because these properties are contextual; their are determined within two different experimental contexts.

We state once again than the essence of QM is not in complementarity, but in contextuality. The real surprise is not that say position and momentum observables are incompatible, but in contextuality (in Bohr's sense) of each of them. The surprise (for classical physicist) is that neither  position nor momentum ``exist'' before measurements, i.e., they cannot be considered as the objective properties of the quantum systems.. 

In the light of Bohr-contextuality, the following natural question arises: 
\medskip

 {\it How can one prove that the concrete observables $a$ and $b$ cannot be jointly measured (i.e., that they are incompatible)?}

\medskip

 From the viewpoint of experimental verification, the notion of incompatibility is difficult. How can one show that the joint measurement of $a$ and $b$ is impossible? One can refer to the mathematical formalism of quantum theory and say that 
 the observables $a$ and $b$ cannot be jointly measurable if the corresponding Hermitian operators $A$ and $B$ do not commute. But, another 
debater can say that may be this is just the artifact of the quantum formalism: yes, the operators do not commute, but observables 
still can be jointly measured. 

\section{Probabilistic Viewpoint on Contextuality-Complementarity}  

The basic analysis on the (in)compatibility problem is done in the probabilistic terms. Suppose that observables $a, b,c,...$ 
can be in principle jointly measured, but we are not able to design the corresponding measurement  procedure. Nevertheless, the assumption of joint measurability, even hypothetical,   implies the existence of JPD. 

\medskip

{\it What are consequences of  JPD's existence?}
 
\medskip
We shall comeback to this question in section \ref{NEE}.  Now we remark that 
the  principle of contextuality-complementarity can be reformulated in probabilistic terms. In short, we can say that {\it the measurement part of QM is a (special) calculus of context-dependent probabilities.} 
This viewpoint was presented in a series of works summarized in monograph \cite{KHR5}  devoted to  the calculus of context dependent 
probability measures $(P_C), C \in \cal{Z},$ where  $\cal{Z}$ is a family of contexts constrained by some consistency conditions. 

We emphasize  that QP  is a special contextual probabilistic calculus. {\it Its specialty 
consists in the possibility to use  a quantum state  (the wave function) $|\psi\rangle$ to unify generally 
incompatible contexts.} This is the important feature of QP playing the crucial role in quantum foundations. 

In classical statistical physics the contextuality of observations is not emphasized. Here it is assumed 
that it is possible to proceed in the CP-framework: to introduce a single  context-independent probability measure $P$ and reproduce the probability distributions of all physical  observables on the basis of $P.$ This is really possible. However, the careful analysis 
of interplay of probability measures appearing in classical physics shows that even here contexuality cannot be ignored. In articles \cite{KHRKOZ1,KHRKOZ2}, there are considered models, e.g., in theory of complex disordered systems (spin glasses), such that it is impossible to operate with just one fixed probability measure $P.$ A variety of context dependent probabilities have to be explored. We especially
emphasize the paper on classical probabilistic entanglement \cite{ALA}.

\subsection{Existence vs. Non-existence  of Joint Probability Distribution}
\label{NEE}

Let  ${\cal P}= (\Omega, {\cal F}, P)$  be a Kolmogorov probability space \cite{K}. Each random variable $a: \Omega \to \mathbf{R}$ determines 
the probability distribution $P_a.$ The crucial point is that all these distributions are encoded in
 the same probability measure $P: P_a (\alpha) = P(\omega \in \Omega: a(\omega) = \alpha).$  (We consider only discrete 
random variables.)

\medskip

{\it In CP,  the probability distributions of all observables (represented by random variables) can be consistently unified on the basis of $P.$}  
\medskip

For any pair of random variables $a, b,$  their JPD $P_{a, b}$ is defined and the following condition of {\it marginal consistency} holds:
\begin{equation}
\label{MC}
P_a(\alpha)= \sum_\beta P_{a,b} (\alpha, \beta)
 \end{equation}
This condition means that observation of $a$  jointly with  $b$ does not change the probability distribution of  $a.$ 
 Equality (\ref{MC}) implies that, for any two observables $b$ and $c,$ 
\begin{equation}
\label{MC1}
\sum_\beta P_{a,b} (\alpha, \beta) = \sum_\gamma P_{a,c} (\alpha, \gamma).
 \end{equation}
 In fact, condition (\ref{MC1}) is equivalent to (\ref{MC}):  by selecting the random variable $c$  such that $c(\omega)=1$ almost everywhere, we see that (\ref{MC1}) implies (\ref{MC}).  These considerations are easily generalized to a system of $k$ random variables $a_1,..., a_k.$ Their JPD  is well defined,
$$
P_{a_1,...,a_k}(\alpha_1,..., \alpha_k)= P(\omega \in \Omega: a_1(\omega) = \alpha_1, ...., a_k(\omega) = \alpha_k).
$$     
And marginal consistency conditions holds for all subsets of random variables $(a_{i_1},...,a_{i_m}), m < k).$

Consider now some system of {\it experimental observables} $a_1,..., a_k.$ If  the experimental design 
for their joint measurement exists, then  it is possible to define their JPD $P_{ a_1,...,  a_k}(\alpha_1,..., \alpha_k)$ (as the relative frequency of their joint outcomes).  This probability measure $P \equiv P_{a_1,...,  a_k}$
can be used to define the Kolmogorov probability space, i.e., the case of joint measurement can be described by CP. 

Now consider the general situation: only some groups of observables can be jointly measured. For example, there are three observables 
$a,b,c$ and only the pairs $(a,b)$ and $(a,c)$ can be measurable, i.e., only JPDs $P_{a, b}$ and $P_{a, c}$ can be defined and associated 
with the experimental data. There is no reason to assume the existence of JPD $P_{a, b,c}.$ In this situation equality (\ref{MC1}) may be violated. In the terminology of QM, this violation  is called {\it signaling.} 

Typically one considers two labs, Alice's and Bob's labs. Alice measures the $a$-observable and Bob can choose whether to measure the $b$- or $c$-observable.  If 
\begin{equation}
\label{MC1mmm}
\sum_\beta P_{a,b} (\alpha, \beta) \not= \sum_\gamma P_{a,c} (\alpha, \gamma),
 \end{equation}
one says that the $a$-measurement procedure is (in some typically unknown way) is disturbed by the selection of a measurement procedure 
by Bob, some signal from Bob's lab approaches Alice's lab and changes the probability distribution.  This terminology, signaling vs. 
no-signaling, is adapted to measurements on spatially separated systems and related to the issue of nonlocality. In quantum-like models, 
one typically works with spatially localized systems and interested in contextuality (what ever it means).  Therefore we called condition  (\ref{MC1}) marginal consistency (consistency of marginal probabilities) and (\ref{MC1mmm}) is marginal inconsistency. In the further presentation we shall use changeably both terminologies, marginal consistency vs. inconsistency and no-signaling vs. signaling.

In future we shall be mainly interested in the CHSH inequality. In this framework, we shall work with  four observables $a_1,  a_2$ and $ b_1,  b_2;$  experimenters are able to design  measurement procedures only  for some pairs of them, say $(a_i, b_j), i, j=1,2.$ In this situation, there is no reason to expect that one can define (even mathematically) the JPD $P_{a_1, a_2, b_1, b_2}(\alpha_1, \alpha_2, \beta_1, \beta_2).$ This situation is typical for QM.   This  is a complex interplay 
of theory and experiment. Only probability distributions $P_{a_i, b_j}$ can be experimentally verified. However, in theoretical speculation, we can consider JPD 
$P_{a_1, a_2, b_1, b_2}$  as {\it mathematical quantity.} If it were existed, we might expect that 
there would be  some 
experimental design for joint measurement of the quadruple of observables $( a_1, a_2, b_1, b_2).$ On the other hand, 
if it does not exist, then it is meaningless even to try to design an experiment for their joint measurement.  

Now we turn back to marginal consistency; in general (if $P_{a_1, a_2, b_1, b_2}$ does not exist), it may be violated. 
However, in QM it is not violated: {\it there is no signaling.} This is the miracle feature of QM. Often it is coupled to spatial 
separation of systems: $a_1$ or $a_2$ are measured on $S_1$ and $b_1$ or $b_2$ on $S_2.$ And these systems are so far from each other that the light signal emitted from Bob's lab cannot approach Alice's lab during the time of the measurement and manipulation with the selection 
of experimental settings. However, as we shall see no-signaling is the general feature of the quantum formalism which has nothing to do with spatial separability nor even with consideration of the compound systems.

\section{ Clauser, Horne, Shimony, and Holt (CHSH) Inequality}

We restrict further  considerations to the CHSH-framework, i.e., we shall not consider other types of Bell inequalities.

How can one get to know whether JPD exists?  The answer to this question is given by a theorem of  Fine \cite{Fine} 
concerning the CHSH inequality.

Consider  dichotomous observables $a_i$ and $b_j (i,j=1,2)$ taking values $\pm 1.$ In each pair $(a_i,b_j)$ observables are compatible, i.e., they can be jointly measurable and pairwise JPDs $P_{a_i, b_j}$ are well defined. Consider correlation    
$$
\langle  a_i  b_j\rangle = E [a_i b_j]= \int  \alpha \beta \;  d P_{a_i, b_j}(\alpha, \beta);
$$ 
in the discrete case,   
$$
\langle  a_i  b_j\rangle = E [a_i b_j]= \sum_{\alpha \beta}  \alpha \beta \;  P_{a_i, b_j}(\alpha, \beta).
$$
By  Fine's theorem  JPD $P_{a_1,a_2,b_1,b_2}$ exists  if and only if the CHSH-inequality for these correlations is satisfied:
\begin{equation}
\label{CH}
\vert \langle  a_1  b_1\rangle  + \langle  a_1  b_2\rangle + \langle  a_2  b_1\rangle - \langle  a_2 b_2\rangle   \vert \leq 2.
\end{equation}
and the three other inequalities corresponding to all possible permutations of indexes $i,j=1,2.$  

\subsection{Derivation of CHSH Inequality within Kolmogorov Theory} 

The crucial assumption for derivation of the CHSH-inequality is that all correlations are w.r.t. the same Kolmogorov probability space 
${\cal P}= (\Omega, {\cal F}, P)$ and that all observables $a_i, b_j, i, j=1,2,$ can be mathematically represented as random variables 
on this space. Under the assumption  of the JPD existence, one can select the sample space 
$\Omega =\{-1, +1\}^4$ and the probability measure  $P=P_{a_1,a_2,b_1,b_2}.$ 
Thus, the CHSH inequality has the form,
\begin{equation}
\label{CH6y}
\Big\vert \int_{\Omega} [a_1(\omega)  b_1 (\omega)  + a_1(\omega)  b_2 (\omega) 
 +  a_2(\omega)  b_1 (\omega)  - a_2(\omega)  b_2 (\omega)] dP(\omega)\Big\vert \leq 2.
\end{equation}
The variable $\omega$ can include hidden variables of a system, measurement devices, detection times, and so on. It is 
only important the possibility to use the same probability space to model all correlations.  The latter is equivalent to the 
existence of JPD $P_{A_1, a_2, b_2,b_2}.$ This is the trivial part of Fine's theorem, JPD implies the CHSH inequality. Another way around is more difficult \cite{Fine}. 

This inequality can be proven by integration of the inequality
$$
- 2 \leq a_1(\omega)  b_1 (\omega)  + a_1(\omega)  b_2 (\omega) +  a_2(\omega)  b_1 (\omega)  - a_2(\omega) b_2 (\omega) \leq 2
$$	
which is the consequence of the inequality
$$
- 2 \leq a_1  b_1 + a_1 b_2 + a_2 b_1- a_2 b_2 \leq 2
$$
which holds for any quadrupole of real numbers belonging $[-1, +1].$ 
		
\subsection{Role of No-signaling in Fine Theorem}	
	
 The above presentation of Fine's result  is common for physics' folklore. However,  Fine did not consider explicitly  the CHSH inequalities presented above, see (\ref{CH}). 
He introduced four inequalities that are necessary and sufficient for 
the JPD to exist, but these inequalities are expressed differently to the CHSH inequalities. 
The  CHSH inequalities are derivable from Fine's four inequalities stated 
in Theorem 3 of his paper.

We remark that the existence of the quadruple JPD implies  marginal consistency  (no-signaling), 
And the Fine theorem presupposed that marginal consistency.  

This is the good place to make the following remark. In quantum physics this very clear and simple meaning of violation of the CHSH-inequality  (non-existence of JPD) is obscured by the issue of  nonlocality.  However, in this book we are not aimed to criticize the nonlocal interpretation of QM.
If some physicists have fun by referring to spooky action at a distance and other mysteries of QM, it is not disturbing for us, since 
we only use the quantum formalism, not its special interpretation.  In any event, non-locality may be relevant only to space separated systems. However, except parapsychology,  cognitive psychology does not handle space separated systems.Finally, we point out  that the Bell type inequalities were considered already by Boole (1862) \cite{Boole1,Boole2} as necessary 
conditions for existence of a JPD.  

\subsection{Violation of CHSH inequality for V\"axj\"o model}

If it is impossible to proceed with the same probability space for all correlations, one has to use the V\"axj\"o model 
(section \ref{VBM}), and there is no reason to expect that the following inequality (and the corresponding permutations) would hold,
\begin{equation}
\label{CH6y1}
\Big\vert \int_{\Omega_{C_{11}}} a_{C_{11}}(\omega)  b_{C_{11}} (\omega) dP_{C_{11}}(\omega) +
\int_{\Omega_{C_{12}}} a_{C_{12}}(\omega)  b_{C_{12}} (\omega) dP_{C_{12}}(\omega)+ 
\end{equation}'
$$
\int_{\Omega_{C_{21}}} a_{C_{21}}(\omega)  b_{C_{21}} (\omega) dP_{C_{21}}(\omega)-
\int_{\Omega_{C_{22}}} a_{C_{22}}(\omega)  b_{C_{22}} (\omega) dP_{C_{22}}(\omega) \Big\vert \leq 2,
$$
where  $C_{ij}$ is the context for the joint measurement of the observables $a_i$ and $b_j.$
Here $a_i$-observable is represented by random variables $(a_{C_{i1}}, a_{C_{i2}})$ and
$b_i$-observable  by random variables $(b_{C_{1i}}, b_{C_{2i}}).$  

In the V\"axj\"o model the condition of no-signaling may be violated; for discrete variables, signaling means that 
$$
\sum_y P_{a_1, b_1|C_{11}} (x,y)\not= \sum_y P_{a_1, b_2|{C_{12}}} (x,y).
$$

\section{CHSH-inequality for quantum observables:  representation via commutators} 
\label{LLL6}

In this section we present the purely quantum treatment of the CHSH inequality and highlight the role of incompatibility in its violation (we follow article \cite{NL1}). Although in QM the CHSH inequality is typically studied for compound systems with the emphasis to the tensor product structure of the state space, in this section we shall not emphasize the latter and proceed for an arbitrary state space and operators. Consequences and simplifications for the tensor product case will be presented in section \ref{LR}.

Observables  $a_i, b_j$ are described by (Hermitian) operators $A_i, B_j, i, j=1,2,$
\begin{equation}
\label{KO1}
[ A_i,  B_j]=0, i,j=1,2.
\end{equation}
We remark that generally
$$
[ A_1,  A_2]\not=0, \; [ B_1,  B_2]\not=0,
$$
i.e., the observables in the pairs $a_1, a_2$ and $b_1, b_2$ do not need to be compatible. 

Observables under consideration are dichotomous with values $\pm 1.$ Hence, the corresponding operators are such that 
$A_i^2= B_j^2 =I.$ The latter plays the crucial role in derivation of the Landau equality (\ref{L2}).

Consider the CHSH correlation represented in the quantum formalism and normalized by 1/2,
\begin{equation}
\label{LC}
\langle  {\cal B} \rangle   =\frac{1}{2} [\langle  A_1  B_1  \rangle + \langle  A_1  B_2 \rangle + \langle  A_2   B_1  \rangle- \langle  A_2  B_2 \rangle].
\end{equation}
This correlation is expressed via \mbox{the Bell-operator:}
\begin{equation}
\label{L1}
 {\cal B} = \frac{1}{2}[ A_1( B_1+  B_2) + A_2( B_1- B_2)]
\end{equation}
as
\begin{equation}
\label{L1T}
\langle  {\cal B} \rangle = \langle \psi\vert  {\cal B} \vert \psi\rangle. 
\end{equation}

Simple calculations lead to the Landau identity \cite{Landau1,Landau2}:
\begin{equation}
\label{L2}
  {{\cal B}}^2= I - (1/4) [ A_1,  A_2][ B_1, B_2].
\end{equation}

If~{\it at least one commutator} equals to zero, i.e.,
\begin{equation}
\label{L3}
[ A_1, A_2]=0, 
\end{equation}
or
\begin{equation}
\label{L4}
[ B_1, B_2]=0,
\end{equation}
then, for quantum observables,  we obtain the inequality
\begin{equation}
\label{L1n}
\vert \langle  {\cal B} \rangle  \vert   \leq 1.
\end{equation}

Derivation of (\ref{L1n}) was based solely on quantum theory. This~inequality is the consequence 
of compatibility for at least one pair of observables, $A_1, A_2$ or $B_1, B_2.$  
Symbolically equation (\ref{L1n}) is the usual CHSH-inequality, but its  meaning is different. 
Equation (\ref{L1n}) can be called  {\it the quantum CHSH inequality.}

Now suppose that $A_i$-observables as well as $B_j$-observables are incompatible, i.e.,~corresponding operators do not commute:
\begin{equation}
\label{L3z}
[ A_1, A_2]\not=0 \; \mbox{and}  \; [ B_1, B_2] \not=0,
\end{equation}
i.e.,
\begin{equation}
\label{L2zT}
 M_A\not=0 \;\mbox{and} \;  M_B\not=0,
\end{equation}
where the commutator observables are defined $M_A= i[ A_1,  A_2], \;   M_B = i [ B_1, B_2].$ We emphasize that  
 $$
[ M_A,  M_B]=0.
$$
The Landau identity can be written as
\begin{equation}
\label{L2z}
  {{\cal B}}^2=I + (1/4) M_{AB},
\end{equation}
where  $M_{AB}=  M_A  M_B= M_B M_A$ is the operator of composition of commutator operators. 

Weremark that if $M_{AB} =0,$ then,   in spite  the incompatibility condition (\ref{L3z}), the quantum QCHSH-inequality cannot be violated. So, we continue under condition
\begin{equation}
\label{L2zz}
 M_{AB}\not=0.
\end{equation}

This condition is not so restrictive. In  my interpretation,  the quantum CHSH-inequality is simply one of possible statistical tests 
of incompatibility. It provides the possibility to estimate the degree of incompatibility in a pair of observables, e.g.,~
in the $A$-pair. The~$B$-pair is the axillary; it can be selected.

The condition in equation~(\ref{L2zz}) is guaranteed via selection of the $B$-operators in such a way 
that the operator     $M_{B}$  is invertible. We point out that the case of compound systems (see section~\ref{LR}) incompatibility of the $A$-observables and the $B$-observables  implies the non-degeneration condition (\ref{L2zz}).

Under condition (\ref{L2zz}), there exists  common eigenvector $\psi_{AB}$ of commuting commutator-operators,  
$$M_A \psi_{AB}= \mu_A \psi_{AB},
M_B \psi_{AB}= \mu_B \psi_{AB}$$ such that  both eigenvalues $ \mu_A,  \mu_B$ are~nonzero. 

Consider the case when $\mu_A>0$ and  $\mu_B >0.$ Such $\psi_{AB}$ is an eigenvector of operator $  {\cal B}^2$ with 
eigenvalue $(1+\mu)>1, \mu=\mu_A \mu_B.$ THus, $\Vert  {\cal B}^2 \Vert \geq (1+ \mu)>1$ and 
$$     
1< (1+ \mu) \leq \Vert  {\cal B}^2 \Vert = \Vert  {\cal B} \Vert^2.
$$

Operator $ {\cal B}$ is Hermitian and this implies that 
$$
\Vert  {\cal B}\Vert = \sup_{\Vert \psi \Vert =1} \vert \langle \psi\vert  {\cal B} \vert \psi\rangle \vert.
$$

Finally,  we obtain the following estimate: 
$$
\sup_{\Vert \psi \Vert =1} \vert \langle \psi\vert  {\cal B} \vert \psi\rangle \vert > \sqrt{1+ \mu} >1.
$$

We demonstrated that, for some pure states,  the quantum  CHSH-inequality f is~violated.

Consider now the case $\mu_A >0,$  but  $\mu_B < 0.$ The sign of   $\mu_B$ can be changed via interchange the $B$-observables.

We conclude:

\medskip

{\it Conjunction of  incompatibilities of  the $A$-observables and the $B$-observables  constrained by equation~(\ref{L2zz}) is sufficient for violation of the quantum CHSH-inequality  (for some quantum state).}

\medskip

The degree of violation can serve as an incompatibility measure in two pairs of quantum observables, $A_1,A_2$ and $B_1, B_2.$ Testing the degree of incompatibility is  testing 
the degree of noncommutativity, or~in other words, the~``magnitudes" of observables corresponding to commutators,
\begin{equation}
\label{INC}
 M_A=i [ A_1,  A_2], \;   M_B=i [ B_1,  B_2].
\end{equation}

The incompatibility-magnitude can be  expressed via the maximal value of averages of commutator-operators, i.e.,~by their norms, for~example,
\begin{equation}
\label{MG}
\sup_{\Vert \psi\Vert=1} \vert\langle \psi\vert  M_A\vert \psi\rangle\vert = \Vert  M_A\Vert.
\end{equation}

By interpreting quantity $\langle \psi\vert  M_A\vert \psi\rangle$ as the 
theoretical counterpart 
of experimental average $\langle M_A\rangle_\psi$ of observable $M_A,$ we can measure  experimentally  the incompatibility-magnitude, i.e.,~norm  $\Vert  M_A\Vert$ 
from measurements of commutator-observable $M_A.$  (The main 
foundational problem is that measurement of such commutator-observables is challenging. Recently some progress was demonstrated on the basis of weak measurements, but~generally we are not able to measure 
commutator-quantities.)  

We remark that (from the quantum mechanical viewpoint) the CHSH-test estimates the product of incompatibility-magnitudes for  the $A$-observables and $B$-observables, i.e.,~the quantity $\Vert  M_A\Vert\Vert  M_B\Vert.$ By~considering 
the $B$-observables as axillary and selecting them in a proper way (for example, such that the $B$-commutator is a simple operator), we can use the CHSH-test to obtain the experimental value for  the incompatibility-magnitude given by $\Vert  M_A\Vert.$

\subsection{Compound~Systems: Incompatibility as Necessary and Sufficient Condition of Violation of Quantum CHSH-Inequality}
\label{LR}

Here, $H=H_A\otimes H_B$ and $   A_j=     {\bf A}_j \otimes I,    B_j=  I \otimes    {\bf B}_j,$
where Hermitian operators   $   {\bf A}_j$ and $   {\bf B}_j$ act in $H_A$ and $H_B,$ respectively. 

Here, the~joint incompatibility-condition in Equation~(\ref{L3z}) is equivalent to incompatibility of observables on subsystems:
\begin{equation}
\label{L3za}
   {\bf M}_A= i [    {\bf  A}_1,   {\bf  A}_2]\not=0 \; \mbox{and}  \;    {\bf M}_B= i [   {\bf  B}_1,   {\bf B}_2] \not=0. 
\end{equation}

We have $   M_{AB}=    M_A    M_B=    {\bf M}_A \otimes    {\bf M}_B.$ 
As     mentioned above, constraint $   M_{AB}\not=0$ is equivalent to (\ref{L3za}). 
Thus,  {\it conjunction of local incompatibilities}
is the sufficient condition for violation of the quantum CHSH-inequality. And we obtain: 

\medskip 

{\bf Theorem 1} [Local incompatibility criteria of CHSH-violation]
 {\it Conjunction of local incompatibilities 
is the necessary and sufficient condition for violation of the quantum CHSH-inequality.} 

\subsection{Tsirelson bound}

By using Landau identity (\ref{L2}) we can derive the {\it Tsirelson bound} $2\sqrt{2}$ for the CHSH correlation of quantum observables, i.e., observables which are represented by Hermitian operators $A_i, B_j, i, j=1,2,$ with spectrum $\pm 1,$
so $A_i^2= B_j^2=I.$ For such operators, for any state $\vert \psi\rangle,$ we have:   
\begin{equation}
\label{CHTs}
\vert \langle  A_1  B_1\rangle  + \langle  A_1  B_2\rangle + \langle  A_2  B_1\rangle - \langle  A_2 B_2\rangle   \vert \leq 2\sqrt{2}.
\end{equation}

On the other hand, if observables are not described by QM,  then this bound can be exceeded. For the V\"axj\"o contextual probability model, the CHSH correlation may approach the value 4. 

\section{Signaling in Physical  and  Psychological Experiments}

By using the quantum calculus of probabilities, it is easy to check whether the no-signaling condition holds for quantum observables, which are represented mathematically by Hermitian operators.  
Therefore Fine's theorem is applicable to quantum observables. This theoretical fact 
played an unfortunate role in hiding from view signaling in experimental research on the violation 
of the CHSH-inequality. Experimenters were focused on observing as high  violation of (\ref{CH}) 
as possible and they ignored the no-signaling condition.
However, if the latter is violated, then a JPD automatically does not exist, and there is no reason to expect 
that  (\ref{CH}) would be satisfied. 
The first paper in which the signaling issue  in  quantum experimental research was highlighted was Adenier and 
Khrennikov (2006) \cite{AD3}. There it was shown that  statistical data collected in 
the basic experiments (for that time) performed by Aspect \cite{ASP} and Weihs \cite{Wei} violates the no-signaling condition. 

After this publication experimenters became aware of the signaling issue and started 
to check it \cite{Gi,Sh}. However, analysis presented in Adenier and 
Khrennikov \cite{AD2}  demonstrated that even statistical data 
generated in the first loophole-free experiment to violate the CHSH-inequality \cite{Hen} 
exhibits very strong signaling.   Nowadays no signaling condition is widely discussed 
in quantum information theory, but without referring to the pioneer works of 
Adenier and Khrennikov \cite{AD3}-\cite{AD2}.

The experiments to check CHSH and other Bell-type inequalities were also  performed for mental observables in the form of questions asked to people \cite{QLX}, \cite{Conte}-\cite{BDZ}. The first such experiment was done in 2008 \cite{Conte} and was based on the theoretical paper of Khrennikov \cite{KHRCOND}.  As was  found by Dzhafarov et al. \cite{DZ3}, all known experiments 
of this type suffer of signaling. Moreover, in contrast to physics, in psychology there are no 
theoretical reasons to expect no-signaling. In this situation Fine's theorem is not applicable. 
And  Dzhafarov and his coauthors were the first who understood  the need of adapting 
 the Bell-type inequalities to experimental data exhibiting signaling.
Obviously, the interplay of whether or not a JPD exists for quadruple 
\begin{equation}
\label{Mx1}
S=(a_1, a_2, b_1, b_2)
\end{equation}
can't be considered  for signaling data.

\subsection {Coupling Method (Contextuality-by-Default)} 
\label{DZH}

Dzhafarov and his coauthors \cite{DZ1}-\cite{DZ5} proposed considering, instead of quadruple $S,$ some octuple ${\bf S}$
generated by doubling each  observable and associating ${\bf S}$ with four contexts of measurements of pairs,
 \begin{equation}
\label{Mx2}
C_{11}=(a_1, b_1), C_{12}=(a_1, b_2), C_{21}=(a_2, b_1), C_{22}=(a_2, b_2).
\end{equation} 
Thus, the basic object of CbD-theory  (for the CHSH inequality) is octuple  of 
observables
\begin{equation}
\label{Mx3a}
(a_{11}, b_{11},  a_{12}, b_{21}, a_{21}, b_{12}, a_{22}, b_{22}),
\end{equation}
so, e.g., observable $a_1$ measured jointly with observable $b_j$ is denoted $a_{1j}.$ 

It is assumed that this system of observables can be realized by random variables on {\it the same Kolmogorov 
probability space} ${\cal P}_{{\bf S}}= ({\bf \Omega}, {\cal \bf{F}},  \bf P).$
We shall use bold symbols for sample spaces and probabilities realizing the octuple representation 
of observables by random variables. 
For example,  $A_{ij}= A_{ij}(\omega), \omega \in\bf  \Omega,$ is a  random variable representing 
observable $a_i$ measured jointly with  the observable $b_j.$  
\begin{equation}
\label{Mx3}
{\bf S}= (A_{11}, B_{11},  A_{12}, B_{21}, A_{21}, B_{12}, A_{22}, B_{22}),
\end{equation}

By moving from quadruple $S$  to octuple ${\bf S},$ one confronts the problem of identity of an observable
which is now represented by two different random variables, e.g., 
the observable $a_i$ is represented by  the random variables $A_{ij}(\omega) , j=1,2.$
In the presence of signaling one cannot expect the equality of two such random variables almost everywhere. 
Dzhafarov et al. came  up with a novel treatment of the observable-identity problem.  

It is assumed that averages
\begin{equation}
\label{Ct}
m_{a; ij}=\langle A_{ij} \rangle, \; m_{b; ij}= \langle B_{ij}\rangle
\end{equation}
and covariation

\begin{equation}
\label{Ct1}
{\cal  C}_{ij} =\langle A_{ij}  B_{ji} \rangle 
\end{equation}
are fixed. These are measurable quantities.  They can be statistically verified by experiment. 

Set 
\begin{equation}
\label{Cty}
\delta(a_i)= m_{a; i1}- m_{a; i2} \; \delta(b_j)= m_{b; j1} - m_{b; j2}, 
\end{equation}
and 
\begin{equation}
\label{Cty1}
\Delta_0 = \frac{1}{2} \Big(\sum_i \delta(a_i) + \sum_j \delta(b_j)\Big).
 \end{equation}
This is the experimentally verifiable measure of signaling.  

We remark that in the coupling representation the joint satisfaction of the CHSH inequalities, i.e.,  (\ref{CH}) and other
inequalities obtained from it via permutations, can be written in the form:
\begin{equation}
\label{CHBJ}
  \max_{ij} \vert \langle  A_{11}  B_{11}\rangle +
\langle  A_{12}  B_{21}\rangle + 
\langle  A_{21}  B_{12}\rangle +
\langle  A_{21}  B_{22}\rangle - 2 \langle  A_{ij}  B_{ji} \rangle\vert  \leq 2.
\end{equation}
In the signaling-free situation, e.g., in quantum physics,  the difference between the left-hand and right-hand sides is considered as 
the measure of contextuality. Denote (1/2 times) this quantity by $\Delta_{\rm{CHSH}}.$ It is also experimentally verifiable.     

Then Dzhafarov and coauthors introduced quantity
\begin{equation}
\label{CHPP0}
\Delta({\bf P})= \sum \Delta_{a_i}({\bf  P})+ \sum \Delta_{b_j}({\bf P}),
\end{equation}
where 
\begin{equation}
\label{CHPP}
\Delta_{a_i}({\bf P})=  
{\bf  P}(\omega: A_{i1}(\omega) \not = A_{i2}(\omega) ), \Delta_{b_j}({\bf P})= 
{\bf P}(\omega: B_{j1}(\omega) \not  = B_{j2}(\omega)).
\end{equation}
Here  $\Delta_{a_i}({\bf P})$ characterizes mismatching of representations of observable $a_i$ 
by  random variables  $A_{i1}$ and $A_{i2}$ with respect to probability measure ${\bf P};$
   $\Delta_{b_j}({\bf  P})$ is interpreted in the same way.
The problem of the identity of  observables is formulated as the mismatching minimization or identity maximization problem   
\begin{equation}
\label{CHPP0a}
\Delta({\bf P}) \to \min
\end{equation}
with respect to all octuple probability distributions ${\bf P}$ satisfying constraints (\ref{Ct}), (\ref{Ct1}). 
And it turns out, that 
\begin{equation}
\label{CHPP0yy}
\Delta_{\rm{min}}=\min \Delta({\bf P}).
\end{equation}
It is natural to consider the solutions of the identity maximization problem (\ref{CHPP0a}) as 
CP-representations for contextual system ${\bf S}.$ 
The corresponding random variables have the highest possible, in the presence of signaling,  degree of identity.  

The quantity 
$$
\Delta_{\rm{min}} - \Delta_0
$$ 
is considered as the measure of ``genuine contextuality''. This approach is very useful to study contextuality in the presence of signaling.      
The key point is the coupling of this measure of contextuality with the problem of the identity of observables measured in different contexts.  As was pointed out in article\cite{DZ4} : 

\medskip

{\it ``...contextuality means that random variables recorded under mutually
incompatible conditions cannot be join together into a single system
of jointly distributed random variables, provided one assumes
that their identity across different conditions changes as little as
possibly allowed by direct cross-influences (equivalently, by observed
deviations from marginal selectivity).''}

This approach to contextuality due to Dzhafarov-Kujala can be reformulated in the CHSH-manner by using what we can call CHSH-BDK inequality:
\begin{equation}
\label{CHBJ}
\max_{ij} \vert \langle  A_{11}  B_{11}\rangle +
\langle  A_{12}  B_{21}\rangle + 
\langle  A_{21}  B_{12}\rangle +
\langle  A_{21}  B_{22}\rangle - 2 \langle  A_{ij}  B_{ji} \rangle \vert  - 2\Delta_0  \leq 2.
\end{equation}
It was proven that  octuple-system ${\bf S}$ exhibits no genuine contextuality, i.e., 
\begin{equation}
\label{CHBJz}
\Delta_{\rm{min}}=\Delta_0,
\end{equation}
  if and only if the CHSH-BDK  inequality is satisfied.
  
\section{Sources of Signaling Compatible with Quantum Formalism}

As was already emphasized, quantum measurement theory is free from signaling: marginals are consistent with JPDs. Now we prove this 
simple fact.

\subsection{Quantum Theory: No-signaling}

Consider the quantum Hilbert space formalism, a state given by density operator $\rho;$ three observables $a,b,c$ represented by operators $A,B,C$ (acting in ${\cal H})$ with spectral families of projectors $E^a(x), E^b(x), E^c(x).$ It is assumed that in each pair $(a, b)$ and  
$(a,c)$  the observables are compatible,  $[A,B] =0, [A, C]=0.$ Then 
\begin{equation}
\label{huhu0}
P(a=x,b=y| \rho) = \rm{Tr} \rho E^a(x) E^b(y), \; P(a=x,c=y| \rho) = \rm{Tr} \rho E^a(x) E^c(y)
\end{equation}
and hence
\begin{equation}
\label{huhu}
\sum_y P(a=x,b=y| \rho)= \rm{Tr} \rho E^a(x) \sum_y E^b(y)= \rm{Tr} \rho E^a(x)
\end{equation}
$$
 = \rm{Tr} \rho E^a(x) \sum_y E^c(y) = \sum_y P(a=x,b=y| \rho).
$$
and we remark that  
\begin{equation}
\label{huhu29}
\rm{Tr} \rho E^a(x) = P(a=x| \rho)
\end{equation}
 and, hence, both marginal probability distributions coincide with the probability of measurement of the $a$-observable alone. 

We remark that this proof of no-signaling can be  easily extended to generalized quantum observables given by POVMs. So, in quantum measurement theory there is no place for signaling. We also recall that signaling (marginal inconsistency) is absent in classical (Kolmogorov) probability theory. On the other hand, it is natural for contextual probability (as in the V\"axj\"o model).     

\subsection{No Signaling for Nonlocal Quantum Observables}

Now let ${\cal H}={\cal H}_1 \otimes{\cal H}_2,$ where ${\cal H}_1, {\cal H}_2$ be the  state spaces of the subsystems $S_1,S_2$ of the compound system $S=(S_1,S_2)$ and let the  observables $a, b, c$ are nonlocal, in the sense that their measurements are not localized to subsystems. The corresponding operators have the form $A= A_1 \otimes A_2, B= B_1 \otimes B_2, C=C_1 \otimes C_2,$ where $A_2, B_1, C_1$ do not need to be equal to $I.$ Let us decompose say $E^a(x)$ into tensor product $E_1^a(x_1) \otimes E_2^a(x_2),$ where outcomes of $a$ are labeled by pairs of numbers $(x_1,x_2) \to x$ (the map from pairs to the $a$-outcomes is not one to one). However, the above general scheme based on (\ref{huhu}) is still valid. The tensor product decomposition of projections does not play any role in summation in (\ref{huhu}). 

\medskip

{\it Nonlocality of observables cannot generate signaling.}

\medskip

This is unexpected fact, because typically signaling is associated with  nonlocality. But, as we have seen, this is not nonlocality of observables. 

Now we turn to the quantum CHSH inequality. As we seen in section \ref{LLL6}, for quantum observables its violation is rigidly coupled  only to their incompatibility.  Even if $A_i=A_{i1}\otimes A_{i2}, i=1,2,$ but $[A_1,A_2]=0,$ then the CHSH inequality is not violated.      

So, by quantum theory signaling is impossible. But, e.g., in  decision making, signaling patterns (expressing marginal inconsistency) were found in all known experiments. This is the contradiction between the quantum-like model for decision making and experiment. This situation questions the whole project on applications of the quantum formalism to modeling behavior of cognitive systems. 

However, there are some ``loopholes'' which can lead to marginal inconsistency. 

\subsection{Signaling on Selection of Experimental Settings}

Consider the Bohm-Bell experiment: a source of photons' pairs $S=(S_1,S_2)$ and two polarization beam splitters (PBSs) in Alice's and Bob's labs; their output channels are coupled to the photo-detectors. Denote orientations of PBSs by $ \theta$ and $ \phi.$ Suppose now that the quantum observables representing measurements on $S_1$ and $S_2$ depend on both orientations, 
\begin{equation}
\label{huhu4}
a=a( \theta,  \phi), b=b( \theta,  \phi).
\end{equation} 
They are represented by operators 
\begin{equation}
\label{huhu4r}
A=A( \theta,  \phi),  B=B( \theta,  \phi).
\end{equation} 
 Thus selection of setting $ \phi$ for PBS in Bob's lab changes the observable (measurement procedure)
in Alice's lab and vice verse. This is a kind of signaling between Bob's lab and Alice's lab, signaling carrying information about selection of experimental settings.\footnote{This can also be referred to the absence of free will of experimenters w.r.t. selection of experimental settings. But, we would not follow this line of thought (which is so natural for philosophy of superdeterminism).} 
In such a situation, 
\begin{equation}
\label{huhu1}
P(a( \theta,  \phi)=x,b( \theta,  \phi)= y) = \rm{Tr} \rho E^{a( \theta,  \phi)}(x) E^{b( \theta,  \phi)}(y)
\end{equation}
and hence
\begin{equation}
\label{huhu2}
\sum_y P(a( \theta,  \phi)=x,b( \theta,  \phi)= y| \rho) = \rm{Tr} \rho E^{a( \theta,  \phi)}(x) \sum_y E^{b( \theta,  \phi)}
\end{equation}
$$
= 
\rm{Tr}  \rho E^{a( \theta,  \phi)} = P(a( \theta,  \phi)=x| \rho),
$$
\begin{equation}
\label{huhu3}
P(a( \theta,  \phi)=x,b( \theta,  \phi^\prime)= y| \rho)= 
\rm{Tr} \rho E^{a( \theta,  \phi^\prime)}(x)   \sum_y E^{a( \theta,  \phi^\prime)}(y) = 
\end{equation}
$$
 \rm{Tr}  \rho E^{a( \theta,  \phi^\prime)}(x) = P(a( \theta,  \phi^\prime)=x| \rho).
$$
Generally 
\begin{equation}
\label{huhu3}
\rm{Tr} \rho E^{a(\theta,  \phi)}(x)\not=  \rm{Tr}  \rho E^{a( \theta,  \phi^\prime)}(x).
\end{equation}
or, in the probabilistic terms, 
\begin{equation}
\label{huhu37}
P(a( \theta,  \phi)=x| \rho) \not= P(a( \theta,  \phi^\prime)=x| \rho).
\end{equation}

We remark that decomposition of $S$ into subsystems $S_1$ and $S_2$ and  association of observables $a$ and $b$ 
with these subsystems did not play any role in quantum calculations. Such decomposition and coupling it with spatial locality 
is important only in the physics as the sufficient condition to prevent signaling on selection of experimental settings.       

In the probabilistic terms each pair of settings determines context $C=( \theta,  \phi)$ and the corresponding 
probability space. Thus, we are in the framework of the V\"axj\"o model for contextual probability. Here 
the possibility of signaling and violation of the Bell type inequalities is not surprising.   

In cognitive experiments, observables are typically questions asked to a system $S$ (e.g., a human). As we have seen,  dependence of questions $a$ and $b$  on the same set of parameters can generate signaling. This dependence is not surprising. Even if questions $a$ and $b$ are processed by different regions of the brain, the physical signaling between these regions cannot be neglected. 
If $\theta$ and  $\phi$ are the contents of the $a$- and $b$-questions, then after a few milliseconds the area of the  brain 
processing $a=a(\theta)$ would get to ``know'' about the content of the $b$-question and thus $a$-processing would depend on both parameter, $a=a(\theta, \phi).$ We remark that an essential part of information processing in the brain is performed via electromagnetic field; such signals propagate with  the light velocity and the brain is very small as a physical body.  

On the other hand, some kind of mental localization must  be taken into account; mental functions performing different tasks use
their own information resources  (may be partially overlapping). Without such mental localization, the brain\footnote{The real situation is more complex; not only the brain, but the whole nervous system is involved in mental processing.} would not be able to discriminate  different mental tasks and their outputs.   At least for some mental tasks (e.g., questions),  dependence of $a$ on the parameter $\phi$ 
(see (\ref{huhu4})) can be weak. For such observables, signaling can be minimized. 

{\it Are there other sources of signaling compatible with quantum formalism?}

\subsection{State Dependence on Experimental Settings}

Let us turn to quantum physics. Here ``signaling'' often has the form of real physical signaling and 
it can reflect the real experimental situation. We now discuss the first Bell-experiment in which the detection loophole was closed \cite{Gi0}. It was  performed in Vienna by Zelinger's group and it was  characterized by statistically significant signaling. By being in Vienna directly after this  experiment, I spoke with people who did it. They told the following story about the origin of signaling - marginal inconsistency. The photon source was based on laser generating emission of the pairs of entangled photons from the crystal. It happened (and it was recognized only afterwards)  that the polarization beam splitters (PBSs) reflected some photons backward and by approaching the laser they changed its functioning and backward flow of photons depended on the orientations of PBSs. In this situation ``signaling'' was not from $b$-PBS to $a$-PBS, but both PBSs sent signals to the source. Selection of the concrete pair of PBSs changed functioning of the source; in the quantum terms this means modification of the state preparation procedure. In this case selection of  a pair of orientations leads to generation of a quantum state depending on this pair, $\rho_{ab}.$ This state modification contributed into the signaling pattern in data. 

The above physical experimental illustration pointed out to state's dependence on experimental context as a possible source of signaling.  It is clear that, for $\rho= \rho_{a,b},$ generally 
\begin{equation}
\label{huhu5}
\rm{Tr} \rho_{ab} E^a(x) \not=  \rm{Tr}  \rho_{ac} E^a(x).
\end{equation}
This dependence also may lead to violation of the Bell inequalities. In the probabilistic terms this is again the area of application of the V\"axj\"o model with contexts associated with quantum states, the probability measures depend on the experimental settings.

We remark that it seems that the state variability depending on experimental settings was the source of signaling in Weihs' experiment
\cite{Wei} which closed nonlocality loophole. At least in this way we interpreted his reply \cite{Wei1} 
to our (me and Guillaume Adenier) paper \cite{AD1}. Since Weihs \cite{Wei} was able to separate two ``labs'' to a long distance, the signals from 
one lab could not approach another during the process of measurement.    

 In quantum physics experimenters were able to block all possible sources of state's dependence on the experimental settings. 
 Thus, it is claimed that one can be sure that $\rho$ does not depend on $a$ and $b.$ By using the orientations of PBSs $\theta, \phi,$ i.e., $\rho=\rho(\theta, \phi),$  the latter condition can be written as 
\begin{equation}
\label{huhu6}
\frac{\partial{\rho(\theta, \phi)}}{\partial \theta}=0, \;  \frac{\partial{\rho(\theta, \phi)}}{\partial \phi}=0.
\end{equation}

Stability of state preparation is the delicate issue. As we have seen, the source by itself can be stable and generate approximately the same state $\rho,$ but the presence of measurement devices can modify its functioning. Moreover, even if any feedback to the source from measuring devices is excluded, laser's functioning can be disturbed by fluctuations. Typically violation of state statsbility cannot be observed directly and the appearance of a signaling pattern can be considered as a sign on state's variation. In physics the 
signaling can be rigidly associated with fluctuations in state preparation. Spatial separation leads to local parameter dependence of observables, i.e., $a=a(\theta)$ and $b=b(\phi).$ 

For cognitive systems, it seems to be impossible to distinguish two sources of signaling:
\begin{itemize}
\item joint dependence on  parameters $\theta, \phi$ determining contents of questions,
\item state dependence on $\theta, \phi.$ 
 \end{itemize}

\section{Nonconetxtual inequalities}
\label{QMC}

As before, we  consider dichotomous observables taking values $\pm 1.$ 

We follow paper \cite{Araujo} (one of the best and clearest representations of noncontextuality inequalities).
Consider a set of observables $\{x_1, . . . ,x_n \};$  contexts $C_{ij}$ determined by   the pairs of  indexes  
such that observables $x_i, x_j$ are compatible, i.e., the pair $(x_i, x_j)$ is jointly measurable; 
set ${\cal Z}= \{C_{ij}\}.$     For  each  context $C_{ij},$  we measure  
correlations   for observables $x_i$ and $x_j$ as well as averages 
$\langle x_i\rangle$ and $\langle x_j\rangle.$ o

The $n$-cycle  contextuality  scenario  is  given  by collection of contexts
\begin{equation}
\label{C}
{\cal Z}_n=\{\ C_{12}, C_{21}, ..., C_{n-1,n}, C_{n 1}\}.
\end{equation}
Statistical data associated with this set of contexts is given by the collection of averages and correlations:
\begin{equation}
\label{Cw}
\{\langle x_1\rangle,...., \langle x_n\rangle; \langle x_1 x_2\rangle, . . . , \langle x_{n-1}x_{n}\rangle, \langle x_{n}x_1\rangle\}.
\end{equation}

Theorem 1 from paper \cite{Araujo} describes all tight noncontextuality inequalities. We are not interested in their general form. 
For $n=4,$ we have inequality:
\begin{equation}
\label{P57}
\vert \langle x_1  x_2  \rangle +  \langle x_2 x_3  \rangle + \langle x_3 x_4 \rangle  - \langle x_4 x_1\rangle \vert \leq 2 .
\end{equation}
This inequality can be rewritten in the QM notation  which we have used in the previous sections 
by setting $x_1=a_1, x_3=a_a, x_2=b_1, x_4=b_2.$

Theorem 2 from article \cite{Araujo} demonstrates that, for $$n \geq 4,$$  aforementioned tight noncontexuality inequalities and, in particular, inequality (\ref{P57}),  are violated by quantum correlations.

\section{Concluding Remarks}

This article is aimed to decouple the Bell tests from the issue of nonlocality via highlighting the contextuality role. We started with discussing the physical meaning of contextuality. The common identification of contextuality with violation of the Bell type inequalities (noncontextual inequalities) cannot be accepted. This situation is illustrated by randomness theory. Here 
the notion of randomness is based on rigorous mathematical formalization. Statistical tests, as e.g. the NIST test, are useful only
to check for randomness the outputs of random or pseudo-random generators. We are also critical to appealing to  JMC and not only because it is based on counterfactuals.  Here it is the good place to recall that Svozil \cite{Svozil1,Svozil2})
and Griffiths  \cite{Griffiths}, \cite{Griffiths1}-\cite{Griffiths3} have the different viewpoint and they suggested experimental tests for JMC. Moreover, Griffiths \cite{Griffiths1} even  claimed that QM is noncontextual. So, the diversity of opinions about ``quantum contextuality'' is really amazing.

Bell considered JMC as an alternative to Einsteinian nonlocality. However, in the framework of the Bohm-Bell experiments, the physical meaning of JMC is even more mysterious than the physical meaning of EPR-nonlocality. JMC gains clear meaning only as the special case of Bohr contextuality. By the latter outcomes of quantum observables are generated in the complex process of the interaction between a system and a measurement apparatus.

Bohr contextuality is the real seed of the complementarity principle leading to the existence of incompatible observables. This principle is also essentially clarified and  demystified through connection with contextuality. Our analysis led to the conclusion that contextuality and complementarity are two supplementary counterparts of one principle. It can be called  the 
{\it contextuality-complementarity principle.}

 This is the good place to mention the studies of Grangier, e.g., \cite{Grangier1,Grangier2}, as an attempt to suggest a heuristically natural interpretation for contextuality, which is different from JMC and Bell contextualities. Grangier contextuality is in fact also closely coupled to the Bohr complementarity principle, although this was not pointed out. 

In the probabilistic terms, Bohr contextuality is represented via the use of a family of Kolmogorov probability spaces which are labeled by experimental contexts. Such formalism, the {\it V\"axj\"o model} for contextual probability.
   
In this review the problem of signaling (marginal inconsistency) is taken very seriously. We (Adenier and Khrennikov) paid attention 
to this problem for many years ago \cite{AD3}-\cite{AD2}. These publications attracted attention of experimenters to signaling problem. Nowadays it is claimed that experimental data does not contain signaling patters. However, our analysis of the first loophole free Bell experiment \cite{Hen} demonstrated that the statistical data suffers of signaling. 

In fact, all data sets which we were  able to get from experimenters and then analyze contain statistically significant signaling patters. By using induction one may guess that even data which
owners claimed no-signaling might suffer of signaling. Unfortunately, I simply do not have resources to lead a new project on data analysis. Moreover, it is still difficult and often not possible at all to receive rough click-by-click data. Creation of the data-base  for all basic quantum foundational experiments is very important for quantum foundations - starting with photo-effect and interference experiments and finalizing with the recent Bell type experiments.

Can one work with statistical data shadowed by signaling? The answer to this question is positive as was shon within recently developed CbD-theory.
It led a new class of inequalities, the Bell-Dzhafarov-Kujala (BDK) inequalities. These inequalities are especially important in quantum-like studies, applications of the quantum formalism outside of physics. Here up to now, all experimental statistical data contains signaling patterns. 

Since incompatibility of quantum observables is mathematically encoded in noncommutativity of corresponding operators, it is natural to 
try to express Bell contextuality with operator commutators.  As was shown in article \cite{NL1}, this is possible at least for the CHSH-inequality.
The basic mathematical result beyond such expression is the Landau inequality \cite{Landau1,Landau2}. In the light of commutator representation of the degree of violation of the CHSH inequality, we suggest to interpret this inequality as a special test of incompatibility of observables.
The commutator representation is valid for any state space, i.e., the tensor product structure does not play any role. In this way we decouple the CHSH inequality from the problem of quantum nonlocality which was so highlighted by Bell. Incompatibility in each pair of local observables and only incompatibility is  responsible for the inequality violation.   

Finally, we study the possible sources of signaling which are not in the direct contradiction with the quantum formalism. One of such sources is disturbance of the state preparation procedure by the selection of the experimental settings. And we discuss  this setting dependent preparations in coupling to the concrete experimental situations. 

\section*{Appendix A: Local realism}

In quantum physics, the violation of the Bell inequalities is coupled to the violation of at least one of the followings two  assumptions:
%\index{ realism} \index{locality} \index{action at a distance}
\begin{itemize}
\item a) realism, 
\item b) locality.
\end{itemize}

Realism is understood as the possibility to assign the values of observables before measurement – to consider the measurement outcome as the objective property of a system. {\it Bohr’s contextuality means violation of the realism assumption.} As was pointed out in section \ref{BPCC}, consideration of such contextuality is meaningful only in the presence of incompatible experimental contexts and hence incompatible observables. From author's viewpoint, the Bell inequality tests are designed to check the existence of incompatible contexts and observables. The violation of these inequalities supports the Bohr complementarity principle and hence  contextuality of quantum observables, i.e., rejection of the realism assumption. From my viewpoint, nothing more can be said about the Bell tests and their foundational implications.

However, Bell highlighted the issue of nonlocality. First, I want to point out to ambiguity of the discussions on ``quantum nonlocality”. Typically, physicists have in mind the violation of Einsteinian locality, a kind of action at a distance \cite{EPR}. Therefore, experimenters separate the subsystems of a compound system as far as possible, to prevent the possibility of communications with light velocity. However, in the derivations of the Bell inequalities the space-time structure does not appear at all. Therefore, 
``Bell (non)locality'' has no direct coupling with Einsteinian (non)locality \cite{KHRV1,KHRV2}.
Note that the difference between the notions of Bell locality, EPR locality, and nonsignaling was first specified
mathematically in article \cite{LE1}. See also \cite{LE2}-- \cite{LE3} for Bell locality and nonlocality.
Bell locality is formulated via the introduction of hidden variables as the factorzation condition, 
see, e.g. \cite{Br}, eq. (3). In fact, Bell nonlocality is a form of JMC expressed in term of hidden variables, 
as Bell pointed out by himself \cite{Bell2}.

This is the good place to remark that by considering the EPR-Bohm correlations in the space-time within the quantum field formalism, one finds that these correlations should decrease with the distance \cite{KHRN1,KHRN2}. The declared conservation of correlations which is apparently confirmed in the Bell experiments is the consequence of the normalization procedure used in these experiments 
\cite{KHRN2}.

\medskip
Now we present some logical considerations:  
\begin{itemize}
\item Local realism = realism and locality
\item Not(Local realism)= Not(realism and locality)= nonrealism or nonlocality,
\end{itemize}
where ``or'' is the non-exclusive or operation. 

The crucial point is that here nonlocality is Bell nonlocality, not Einsteinian one.  Hence, nonlocality = JMC (expressed with hidden variables).  And it is a consequence of Bohr contextuality; this can also be said  about nonrealism.

Thus, the whole Bell consideration can be reduced to showing that by rejecting the Bohr contextuality-complementarity principle  one can derive special inequalities for correlations.  From my viewpoint, the violation of these inequalities implies only that the Bohr principles hold true. Roughly speaking one can come back to the foundations of QM which were set 1920th. The experimental Bell tests are advanced tests of the Bohr contextuality-complementarity principle; in this  sense they are tests of quantumness.

 We remark that original Bohr and Heisenberg appealing to the Heisenberg uncertainty relation as the basic test of incompatibility
for quantum observables, e.g.,  \cite{BR0}, \cite{BR1}--\cite{BR2a}was strongly criticized, e.g., by Margenau \cite{Margenau} and {Ballentine \cite{BL1,BL2}. Since direct measurement of the commutator observable  $C= i[A, B]$ is difficult, the Bell tests became the most popular tests of incompatibility
and, hence, quantumness. Unfortunately, the issue of incompatibility was shadowed by ``quantum nonlocality''. 

\section*{Appendix B: Kolmogorov axiomatization of probability} 
 
The {\it Kolmogorov probability space} \cite{K} is any triple of the form
$$(\Omega, {\cal F}, P),$$ 
where $\Omega$ is a set of any origin and ${\cal F}$ is a
$\sigma$-algebra of its subsets, $P$ is a probability measure on ${\cal F}.$

The set $\Omega$ represents random parameters of the model.
In mathematical literature the elements of $\Omega$ are called {\it elementary events}.  {\it Events} are special sets of elementary events, those belonging to the $\sigma$-algebra  ${\cal F}.$

We remind that a $\sigma$-algebra is a set-system
containing $\Omega$ and empty set and closed w.r.t. countable unions and intersections and complements. 

For example, the collection of all subsets of $\Omega$ is a $\sigma$-algebra. This $\sigma$-algebra is used in the case of finite or countable set $\Omega,$
\begin{equation}
\label{NORM27}
\Omega=\{\omega_1,...,\omega_n,...\}.
\end{equation} 

The probability is defined as a {\it measure}, i.e., a map from ${\cal F}$ to non-negative real numbers which is $\sigma$-additive:
\begin{equation}
\label{NORM2}
P(\cup_{j} A_j) = \sum_j P(A_j),
\end{equation}
where $A_j \in {\cal F}$ and $A_i\cap A_j= \emptyset, i \not=j.$ The probability measure is always normalized by one:
\begin{equation}
\label{NORM}
P(\Omega)=1.
\end{equation}

In the case of a discrete probability space, see (\ref{NORM27}), the probability measures have the form
$$
P (A) = \sum_{\omega_j \in A} p_j, \; p_j=P(\{\omega_j\}).
$$

\end{document}